# Calorimetric Power Measurements in the EAST ECRH System

Weiye Xu,[a)] Handong Xu, Fukun Liu, Jian Wang, Xiaojie Wang, Yongzhong Hou

*Institute of plasma physics, Chinese academy of sciences, Hefei 230031, China*

**Abstract:** In this paper, the calorimetric power measurement method for electron cyclotron resonance heating system on EAST are presented. This method requires measurement of the water flow through the cooling circuits and the input and output water temperatures in each cooling circuit. Usually, the inlet water temperature is controlled to be stable to get more accurate results. The influence of the inlet water temperature change on the measurement results is analyzed for the first time in this paper. A novel temperature calibration method is proposed also. This kind of calibration method is accurate and effective, and can be easily implemented.

## I. INTRODUCTION

A 140GHz electron cyclotron resonance heating system for EAST (Experimental Advanced Superconducting Tokamak) is being built in ASIPP (Institute of Plasma Physics, Chinese Academy of Sciences)[1]. This project is designed to inject 4MW/140GHz/100s RF power to EAST. The first two gyrotrons have been established already. The two gyrotrons are produced by Gycom and CPI respectively, and they have been tested in ASIPP. To get an accurate interpretation of the energy balance of the plasma and its response to heating during electron cyclotron resonance heating (ECRH) experiments, a reliable estimate of the RF power delivered to the plasma is needed[2].

There are many methods to measure the millimeter wave power, such as calorimetric method (which may use thermistor sensors, thermoelectric transducers[3], etc.), detection method (which may use diode sensors), etc. These methods which use thermistor sensors, thermoelectric sensors and diode sensors can only measure small power. They can not be used to measure the output power of the gyrotrons directly because the output power is in MW-level. We can use the couplers to couple small amounts of microwave energy from the high power to measure the real power. In addition, we can make big dummy loads which are cooling by water flow for gyrotrons, then we can measure the output power of the gyrotrons by measuring the calories of the dummy load. This method is called the flow calorimetric method, or calorimetric method for short in this paper. The calorimetric method is used to measure the output power

____________________

Electronic mail: xuweiye@ipp.cas.cn

of gyrotrons in EAST ECRH system. This method is accurate only when the inlet water temperature is stable. The influence of the inlet water temperature change on the measurement results is analyzed for the first time in this paper.

In this paper, we begin by discussing the calorimetric power measurements. Next, we analyze the calorimetric method when the inlet water temperature is variable. Then, some test results are given. Finally, we give the conclusion.

## II. CALORIMETRIC POWER MEASUREMENTS

The calorimetric measurements require the measurements of the water flow through each cooling circuit and the input and output water temperature in each cooled circuit. We have established a stable data acquisition system to acquire the water flows and the temperatures in each circuit[4]. It is based on the PXI platform. PXI is a rugged PC-based platform for measurement and automation systems [5]. It is widely used in the area of measurement[6, 7], data acquisition[8], control[9, 10], etc. In our system, the data is being acquired when the gyrotron is working, being displayed on the monitor in real time, and being saved to the data storage server which are ready to be used for successive analysis procedures.

The architecture of the calorimetric measurement is presented in Fig. 1. In EAST ECRH system, temperatures are measured using platinum resistance temperature transmitters to an accuracy of about $\pm$ 0.3℃ (0℃ to 100℃) and a resolution of 0.02℃, water flow is measured using vortex meters to an accuracy of about 1% F.S.. These measured data then be send to PXI 6229 data acquisition cards [11] through the isolators. PXI 6229 DAQ cards have an accuracy of 3.1mV when the range is -10V to 10 V, and having a sensitivity of 97.6μV when the range is -10V to 10V. The isolators have an accuracy of 10mV. So, when we set the range of DAQ cards to be -10V to 10V, the voltages from 0V to 10V are proportional to the temperatures from 0℃ to 100℃. So, the relative error of temperature measurement is,

$$\left|\frac{\Delta T}{T}\right| = \sqrt{\left(\left|\frac{0.3}{T}\right|^2 + \left|\frac{0.1}{T}\right|^2 + \left|\frac{0.031}{T}\right|^2\right)} \approx \left|\frac{0.3177}{T}\right| \qquad (1)$$

where $\left|\frac{0.3}{T}\right|$ is caused by temperature transmitters, $\left|\frac{0.01}{T/10}\right| = \left|\frac{0.1}{T}\right|$ is caused by the isolators, $\left|\frac{0.0031}{T/10}\right| = \left|\frac{0.031}{T}\right|$ is caused by the DAQ cards.

Similarly, the relative error of flow measurement is,

$$\left|\frac{\Delta F}{F}\right| = \sqrt{\left(\left|\frac{0.01 F_{full}}{F}\right|^2 + \left|\frac{0.001 F_{full}}{F}\right|^2 + \left|\frac{0.00031 F_{full}}{F}\right|^2\right)} \approx \left|\frac{F_{full}}{100 F}\right| \quad (2)$$

where $F_{full}$ is the preset maximum flow. For instance, $F_{full}$ is set as 100 m³/h for Gycom Dummy load.

The microwave output power generated by a specific gyrotron can be measured by directing the beam to dummy load which is cooled by deionized water. By measuring the inlet water temperatures and the outlet water temperatures of dummy load, we can calculate the power absorbed by dummy load [2, 12, 13]. We can use the same way to get the power absorbed by main window[14], MOU, waveguides, waveguide switch, miter bends, and bellows which can be used to estimate the gyrotron output power when the rf power is directed to tokamak. The generated power of gyrotron is the sum of those power data mentioned above. We also use the same way to get the power absorbed by relief load, and relief window.

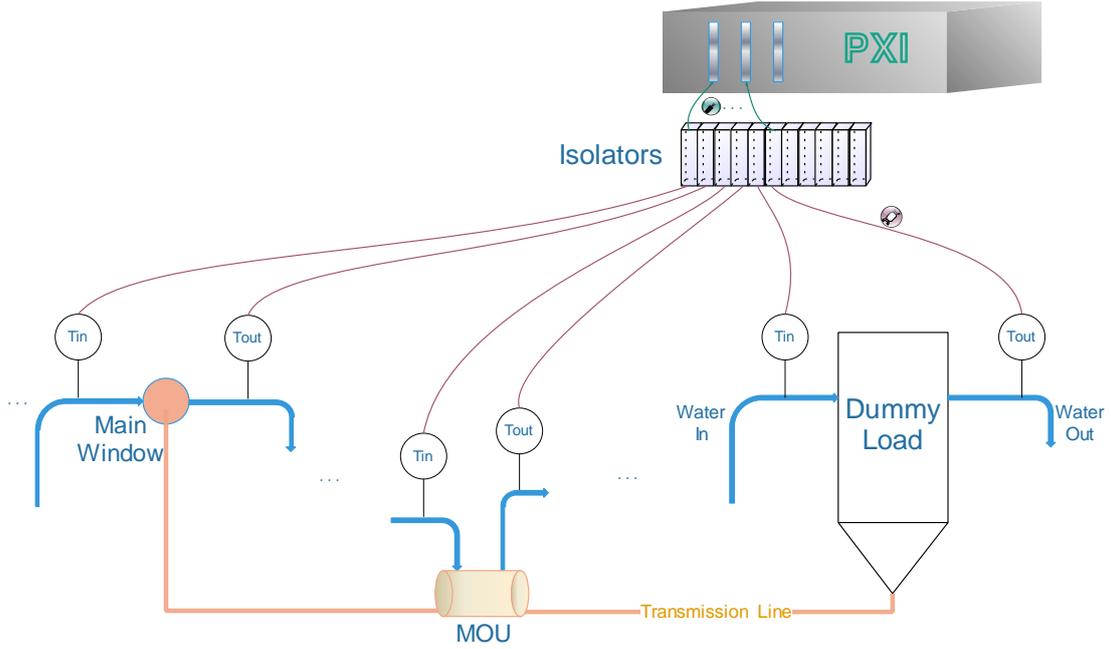

FIG. 1. Architecture of calorimetric measurements.

As shown in Fig. 2, $t_0$ is the initial time. The water, whose temperature is $Tin(t_0)$, run through cross section A at time $t_0$ to point B after time δt. The temperature of the water is $Tout(t_0 + δt)$ at the point of B.

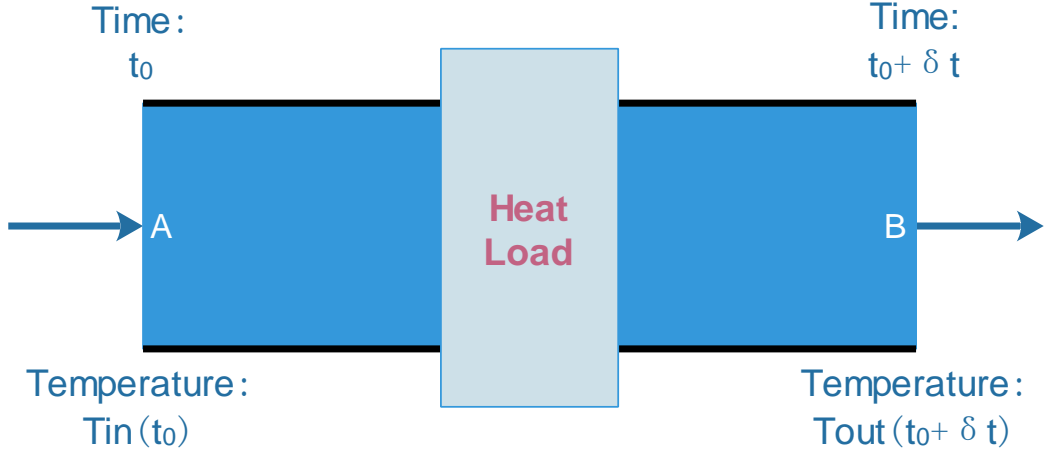

FIG .2. Schematic of calorimetric measurements.

We know that the energy rise of the water can be calculated from

$$W = C \cdot m \cdot \Delta T \qquad (3)$$

where C is the specific heat capacity of water, m is the mass of the water, $\Delta T$ is the temperature rise. So, from time $t_0 + \delta t$ to time $t_0 + \delta t + \Delta t$ at point B, the energy rise from point A to point B is,

$$W_{t0+\delta t+\Delta t} = C \cdot m \cdot [Tout(t_0 + \delta t + \Delta t) - Tin(t_0)] \qquad (4)$$

where $m = F \cdot \Delta t$, $F$ is the mass flow of water in this circuit, it is controlled to be stable in experiments. So, $F$ can be seen as a constant here. Similarly, from time $t_0$ to time $t_0 + \Delta t$ at point B, the energy rise is,

$$W_0 = C \cdot m \cdot [Tout(t_0 + \Delta t) - Tin(t_0 - \delta t)] \qquad (5)$$

So, from time $t_0$ to time $t_n$, all energy rise from point A to point B is,

$$W = \sum_{i=0}^{n} W_i = \sum_{i=0}^{n} C \cdot F \cdot \Delta t \cdot [Tout(t_i + \Delta t) - Tin(t_i - \delta t)] \qquad (6)$$

where $t_{i+1} - t_i = \Delta t$.

When $\Delta t \to 0$,

$$W = \lim_{\Delta t \to 0} \sum_{i=0}^{n} C \cdot F \cdot \Delta t \cdot [Tout(t_i + \Delta t) - Tin(t_i - \delta t)] = \lim_{\Delta t \to 0} \sum_{i=0}^{n} C \cdot F \cdot \Delta t \cdot [Tout(t_i) - Tin(t_i - \delta t)] = C \cdot F \cdot \int_{t_0}^{t_n} [Tout(t) - Tin(t - \delta t)] dt \qquad (7)$$

Assume the wave-output duration is $\tau$, the average power of the gyrotron over the time span $\tau$ can be got by

$$P = \frac{W}{\tau} = \frac{C \cdot F}{\tau} \cdot \int_{t_1}^{t_n} [Tout(t) - Tin(t - \delta t)] dt \qquad (8)$$

We found that when gyrotron ceases generating rf power, the difference between outlet water temperature and inlet water temperature rises still, it comes to the maximal value after a while, then it descends slowly to zero within 60 seconds after the pulse is ended, as shown in Fig. 3. This means that energy is absorbed and temporarily stored in structural material with possible thermal loss to the outside. The thermal loss is assumed to be little enough to be ignored. So,

$$P = \frac{C \cdot F}{\tau} \cdot \int_{t_0}^{+\infty}[Tout(t) - Tin(t - \delta t)]dt \approx \frac{C \cdot F}{\tau} \cdot \int_{t_0}^{t_n=t_0+\tau+60}[Tout(t) - Tin(t - \delta t)]dt \quad (9)$$

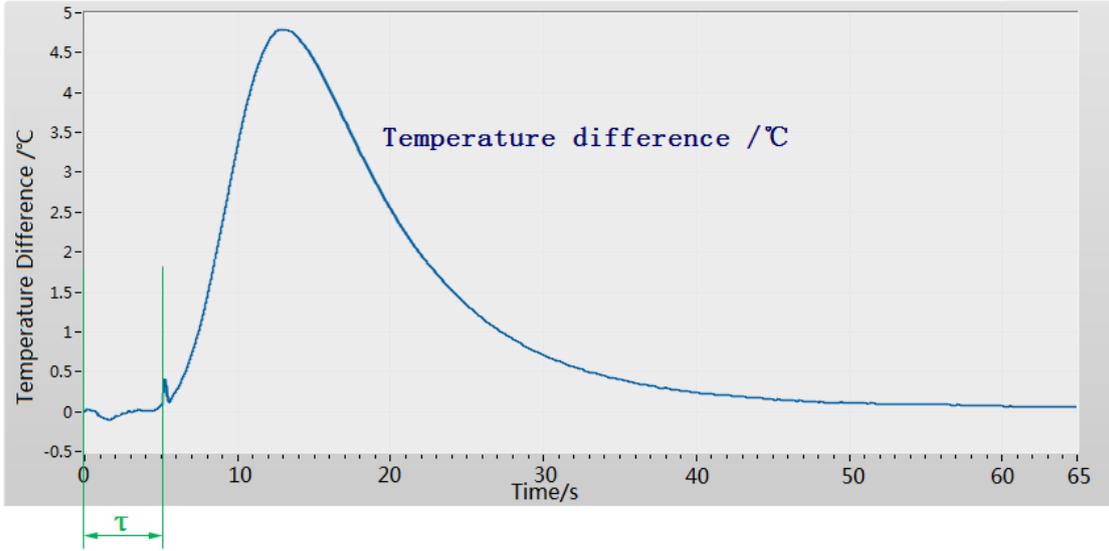

FIG. 3. The typical signal of temperature difference between output water and input water. The pulse duration is $\tau$.

In the measurement, the water temperature at the inlet is kept as an approximate constant, thus, formula (9) can be simplified to,

$$P = \frac{C \cdot F}{\tau} \cdot \int_{t_0}^{t_n=t_0+\tau+60}[Tout(t) - \text{Tin}]dt = \frac{C \cdot F}{\tau} \cdot \int_{t_0}^{t_n=t_0+\tau+60}[Tout(t) - Tin(t)]dt \quad (10)$$

As mentioned before, the absolute error of the measured temperature is about $\pm 0.3177°C$, so, the maximum power error is

$$P_{error} = \frac{C \cdot F}{\tau} \cdot \int_{t_0}^{t_n=t_0+\tau+60} \pm 0.6354\, dt = \pm \frac{C \cdot F}{\tau} \cdot 0.6354 \cdot (\tau + 60) \quad (11)$$

But when the pulse is very short, the maximum of $Tout(t) - Tin(t)$ is smaller than 0.6354 °C in some circuits. So, the real power,

$$P_{real} = \frac{C \cdot F}{\tau} \cdot \int_{t_0}^{t_n=t_0+\tau+60}[Tout(t) - Tin(t)]dt < \frac{C \cdot F}{\tau} \cdot 0.6354 \cdot (\tau + 60) \quad (12)$$

That is, the relative error of the power can be up to $\pm 100\%$. In order to solve this problem, we must calibrate the temperature difference first. The real temperature difference equals the measured temperature difference plus a base temperature, an example is shown in Fig. 4.

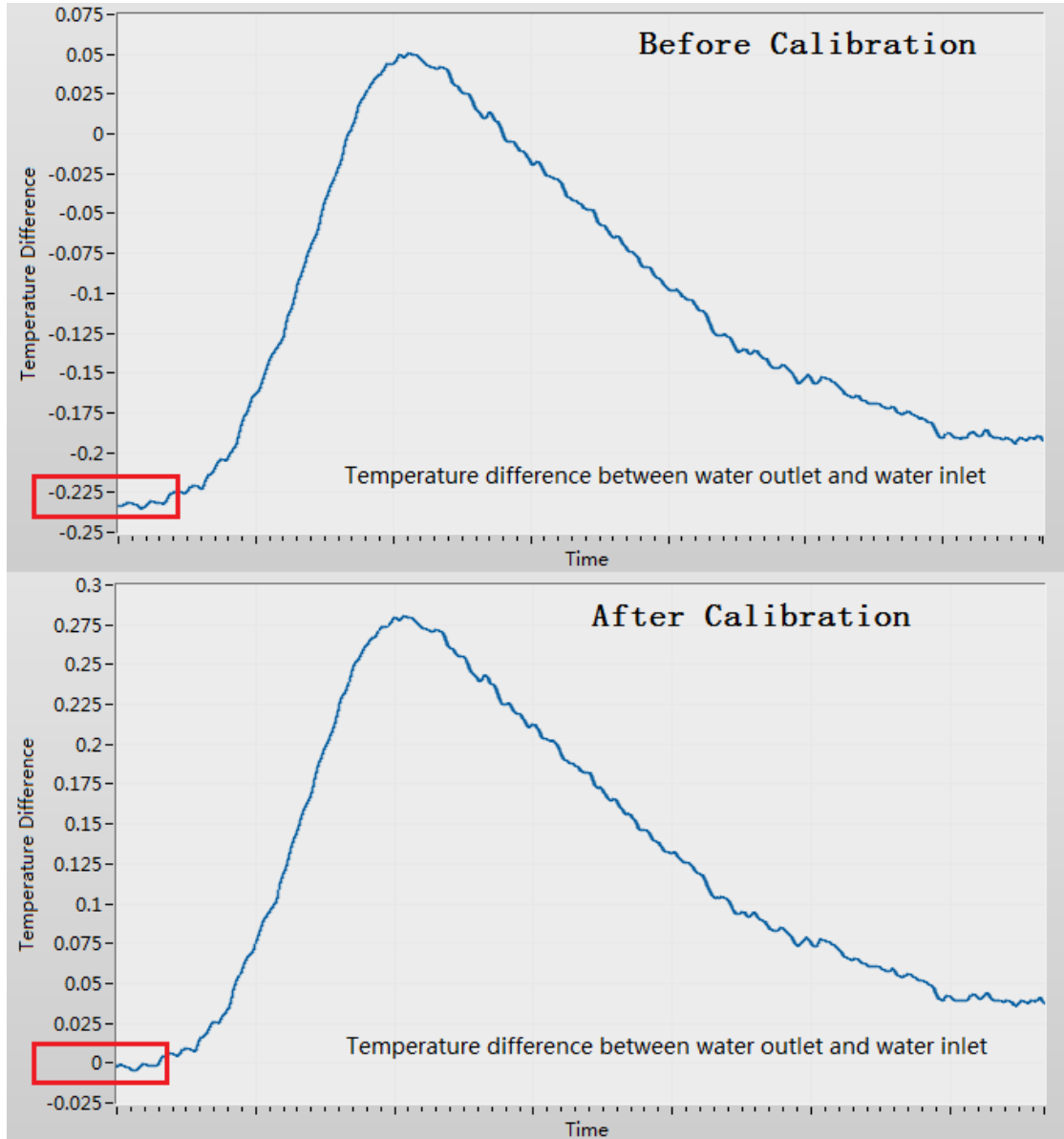

FIG. 4. The temperature is influenced by calibration. The relative temperature is more accurate than the absolute temperature, it is influenced by the resolution of the temperature sensor only.

In the case of no RF power transmitted to the load, the inlet water temperature should be equal to the outlet water temperature. But due to the existence of the measurement error of the temperature sensors, the difference between the inlet temperature and the outlet temperature is not zero but an approximate fixed value. So we need to calibrate the temperature measurement to eliminate the error. That is, we need to find the difference value between the outlet water temperature and the inlet water temperature in the absence of RF power.

There are two methods to get the compensation value. One method is using the digital low pass filter to filter out the noise signals of the temperature signals. Then we can get the inlet temperature $T_{in}$ and outlet temperature $T_{out}$. So, the compensation value is,

$$T_c = T_{in} - T_{out} \tag{13}$$

In the ideal case, the temperature difference after calibration becomes,

$$\Delta T_{ac} = T_{out} - T_{in} + T_c = 0 \tag{14}$$

So, ideally, when no RF power is transmitted to dummy load, the measured power will be approximately equal to zero. In fact, the ambient temperature is not a constant. Even if the water temperature is completely constant, the temperature measured by sensors will always be a little jitter. So, the compensation value is not particularly accurate using the method of low pass filter.

The other method to get the compensation value is using integral formula (10). We give a test pulse signal with duration τ (taking into account the time measurement error, 10s>τ>500ms is appropriate) to start the power integral program. Because no RF power transmitted to dummy load, the measured power should be zero. But due to the temperature difference between the outlet water and the inlet water, the measured power using formula (10) is not zero, but is,

$$P_{cal} = \frac{C \cdot F}{\tau} \cdot \int_{t_0}^{t_n = t_0 + \tau + 60} [Tout(t) - Tin(t)]dt = \frac{C \cdot F}{\tau} \cdot \Delta T \cdot (\tau + 60) \tag{15}$$

Where, $\Delta T$ is the average temperature difference between the inlet and the outlet. An example of the waveforms of the calorimetric power measurements in the absence of RF power is shown in Fig. 5. The test pulse duration is 0.896s. As we can see, the measured dummy load power is gradually increasing with an approximately constant slope. Using the measured power value, the compensation value can be calculated by,

$$T_c = -\Delta T = -\frac{P_{cal}\tau}{CF(\tau+60)} \tag{16}$$

All variables in formula (16) are in the International System of Units. If we change the unit of flow signal $F$ from 'kg/s' to the commonly used unit 'm³/h', the compensation value can be calculated by,

$$T_c = -\Delta T = -\frac{3.6\, P_{cal}\tau}{CF(\tau+60)} \tag{17}$$

This method is equivalent to find the average temperature difference between the inlet and the outlet. This method is more accurate than the first method of low pass filter. And it is easy to implement because this calibration method is using power measurement principle to calculate the compensation value. We choose this approach to calibrate the temperature measurement. As the ambient temperature and water flow are not constant, we need to calibrate the calorimetric power measurement system periodically to calculate the latest calibration value to get more accurate measurement results.

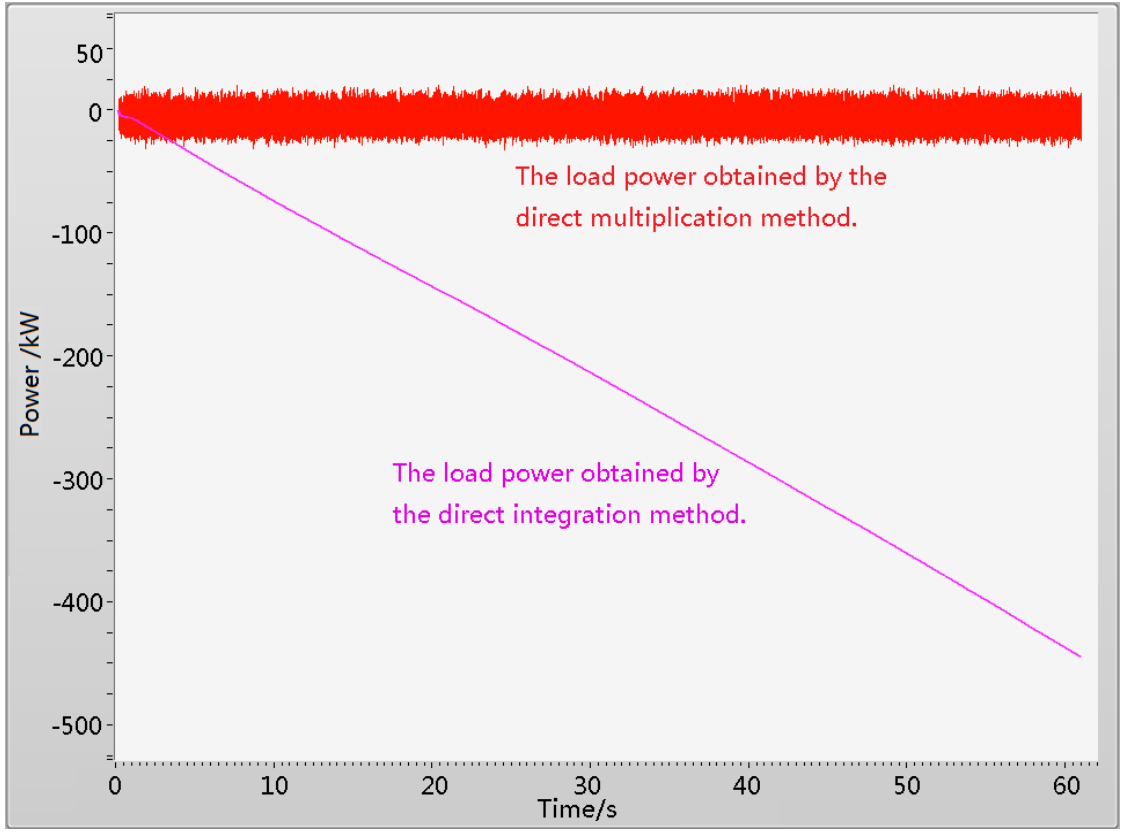

FIG. 5. An example of the waveforms of the calorimetric power measurements of the dummy load. The dummy load power was measured in the absence of RF power.

The temperature difference error is effectively corrected by calibration. But the resolution of temperature transmitter can cause measurement error also. Actually, after temperature calibration, the power measurement error is caused mostly by the resolution of temperature transmitter and the accuracy of the flow meter. Assume,

$$T_{int} = \int_{t_0}^{t_n=t_0+\tau+60}[Tout(t) - Tin(t)]dt \quad (18)$$

Then, we have,

$$P_{real} = \frac{C \cdot F}{\tau} \cdot T_{int} \quad (19)$$

In the worst case, the relative error of measured power is,

$$\delta_{power} \approx \pm \left[\left|\frac{\partial \ln(P_{real})}{\partial F}\Delta F\right| + \left|\frac{\partial \ln(P_{real})}{\partial T_{int}}\Delta T_{int}\right| + \left|\frac{\partial \ln(P_{real})}{\partial \tau}\Delta \tau\right| + \left|\frac{\partial \ln(P_{real})}{\partial C}\Delta C\right|\right] = \pm \left(\left|\frac{\Delta F}{F}\right| + \left|\frac{\Delta T_{int}}{T_{int}}\right| + \left|\frac{\Delta \tau}{\tau}\right| + \left|\frac{\Delta C}{C}\right|\right) \quad (20)$$

where $\Delta F$, $\Delta T_{int}$, $\Delta \tau$ is the absolute error of flow, temperature integral, wave-output duration respectively. In our system, as see in formula (2),

$$\left|\frac{\Delta F}{F}\right| \approx \left|\frac{F_{full}}{100F}\right| \tag{21}$$

In our system, the duration measurement error is about 1ms, and less than 2ms in the worst case,

$$\Delta \tau < 2ms = 0.002s \tag{22}$$

We set the specific heat capacity of water as,

$$C = 4.18 \text{ kJ}/(\text{kg} \cdot °C) \tag{23}$$

But actually, the specific heat capacity will change along with the temperature and the pressure, as shown in Table 1. And the water temperature is between 0 and 100 °C and the water pressure is between 0.1 MPa and 1 MPa. So, the maximum relative error of the specific heat capacity is,

$$\left|\frac{\Delta C}{C}\right| = \frac{4.217 - 4.18}{4.217} \times 100\% \approx 0.88\% \tag{24}$$

TABLE 1. Liquid water specific heat capacity at constant pressure along with the temperature and the pressure. The unit is kJ/(kg · °C).

| Pressure (MPa) | Temperature (°C) | | | |
|---|---|---|---|---|
| | 0 | 20 | 50 | 100 |
| 0.1 | 4.217 | 4.182 | 4.181 | |
| 0.5 | 4.215 | 4.181 | 4.180 | 4.215 |
| 1.0 | 4.212 | 4.179 | 4.179 | 4.214 |

The typical temperature difference curves are shown in Fig. 6. Assume $T_{max}$ is the maximum of the difference between final and initial water temperatures in each circuit, $t_r$ is the duration when temperature difference rise from zero to maximum (It is related to the response time of temperature transmitter, the character of load, and the gyrotron pulse duration). As we can see,

$$T_{int} \approx \max(\tau, t_r) \cdot T_{max} \tag{25}$$

Because the temperature resolution error is $\pm 0.02$ °C, in the worst case,

$$\Delta T_{int} \approx 0.04 \cdot (\tau + 60) \tag{26}$$

Hence,

$$\left|\frac{\Delta T_{int}}{T_{int}}\right| \approx \frac{0.04 \cdot (\tau + 60)}{\max(\tau, t_r) \cdot T_{max}} \tag{27}$$

But due to the randomness of the temperature resolution error, usually, we have,

$$\Delta T_{int} \approx 0 \tag{28}$$

So, the maximum calorimetric measurement error is,

$$\delta_{power} \approx \pm \left(\frac{F_{full}}{F} + \frac{4 \cdot (\tau + 60)}{\max(\tau, t_r) \cdot T_{max}} + \frac{0.2}{\tau} + 0.88\right)\% \tag{29}$$

The probable calorimetric measurement error is,

$$\delta_{power} \approx \pm \left(\frac{F_{full}}{F} + \frac{0.2}{\tau} + 0.88\right)\% \tag{30}$$

where $F$ is the mass flow of water in each circuit; $T_{max}$ is the maximum of the difference between final and initial water temperature in each circuit; $\tau$ is wave-output duration; $t_r$ is the duration when temperature difference rise from zero to maximum, for instance, $t_r \approx 3$ s for waveguide dummy load of CPI gyrotron when the pulse duration is 2 s, $t_r \approx 14$ s for aluminum tank dummy load of CPI gyrotron when the pulse duration is 2 s, $t_r \approx 13$ s for dummy load of Gycom gyrotron when the pulse duration is 5 s. And all parameters are in standard international unit. When the gyrotron pulse duration is bigger than 1 s, the relative error of calorimetric measurement is less than 10% and less than 5% in most cases.

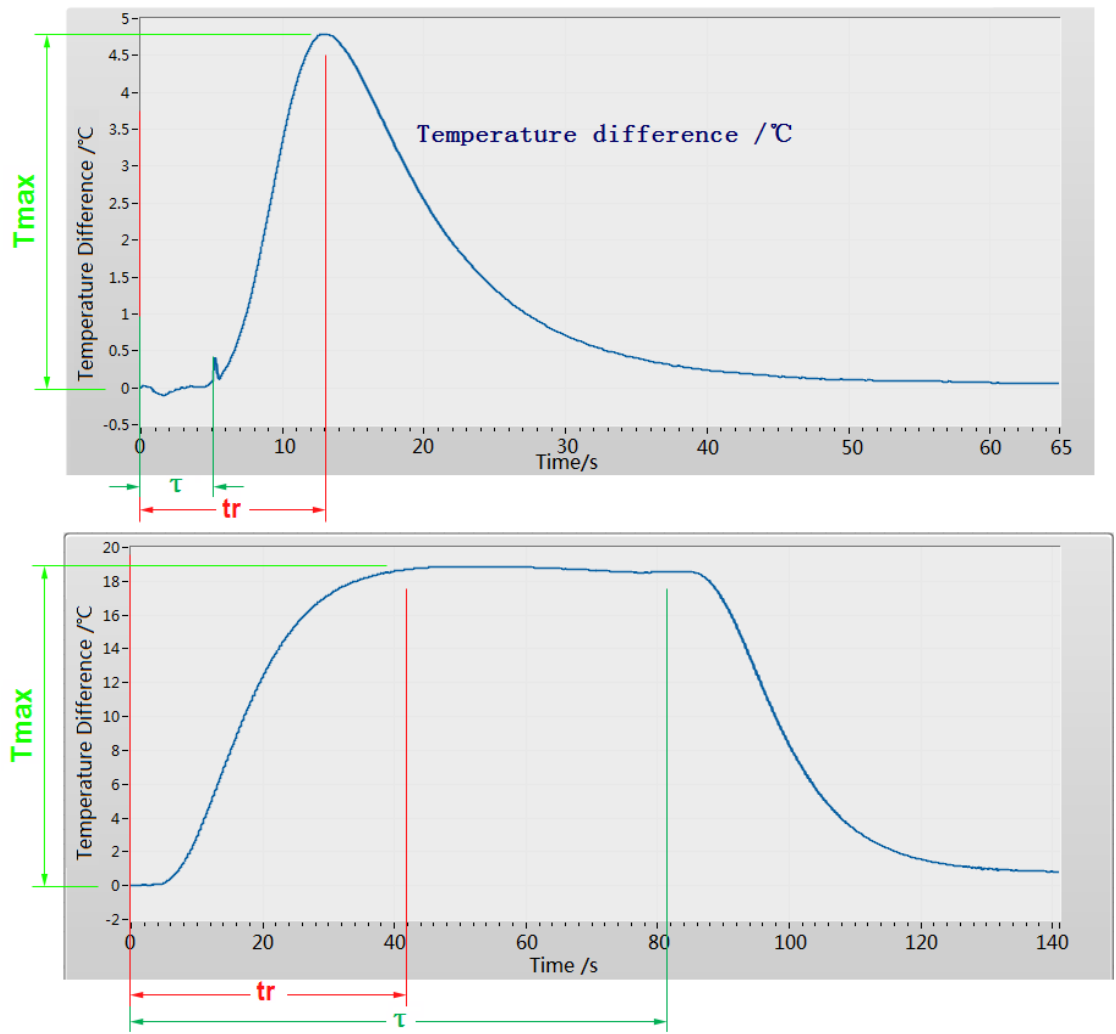

FIG. 6. The temperature difference between outlet and inlet of dummy load cooling circuit. The figure above shows the situation when tr is bigger than τ. The figure below shows the situation when tr is smaller than τ.

## III. CALORIMETRIC MEASUREMENTS WHEN THE INLET WATER TEMPERATURE IS VARIABLE

When the water temperature at the inlet is variable, the power measurement can be got by formula (9). Firstly, the parameter δt should be measured. Assume the water flow is F, the water volume of the dummy load is V, then, we have,

$$\delta t = \frac{V}{F} \qquad (31)$$

where F is volume flow in the unit of m$^3$/h. We can get the volume V by injecting water to the empty dummy load, assume the volume flow of water is $F_0$, the spend time from beginning to inject the water to the time when the outlet flow out water is $\delta t_0$. So,

$$V = F_0 \cdot \delta t_0 \qquad (32)$$

$$\delta t = \frac{F_0 \cdot \delta t_0}{F} \qquad (33)$$

In theory, the power can be measured by formula (9) and (33) when the inlet water temperature is variable. But in fact, the real-time measurement of the water temperature is difficult. In our system, the maximum response time of the temperature transmitter is about 18s. So, this measurement method is not so accurate because the temperature measurement delay. Usually, we measure the power by keeping the inlet water temperature as a constant. But for verifying this measurement theory is correct, we do some experiments.

As shown in Fig. 7, the inlet water temperature increases gradually at a constant slope. We can get the correct power using formula (9). The definite integral is calculated as the green rectangle shown. The temperature difference at time t should be the outlet water temperature at time t minus the inlet water temperature at time *t-δt*. But if we use formula (10) to calculate the power, the definite integral is calculated as the brown rectangle shown, the temperature difference is *ΔT* less than the correct value in every moment. So, the measurement result using formula (10) will be smaller than the correct result.

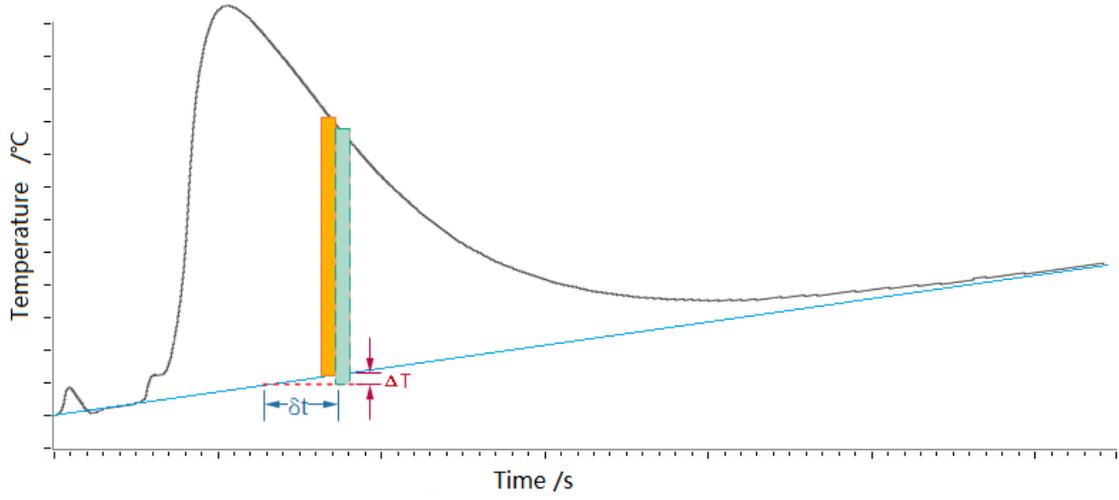

FIG. 7. The temperature of the outlet water and the inlet water along with the time when the inlet water temperature is gradually increasing. The black curve indicates the outlet water temperature and the blue curve indicates the inlet water temperature.

We make the inlet water temperature increase gradually in the gyrotron pulse, and use formula (10) to calculate the power. Some test results are shown in Table 2. Especially, shot number 1 is measured when the inlet water temperature is gradually increasing in the pulse duration. Obviously, the power test results (506kW, 812 kW) are smaller than the real value (546 kW, 849 kW). The temperature difference of the dummy load circuit is shown in Fig. 8. The inlet water temperature rose 0.09 ℃ after 60 s.

When we use formula (10) to calculate the power, the temperature difference is $\Delta T$ less than the correct value in every moment, and,

$$\delta t = \frac{\Delta T}{0.09} \times 60 \tag{34}$$

By formula (10), we can get,

$$\frac{F_V \cdot C \cdot \Delta T \cdot 60}{3.6 \cdot \tau} = \frac{42 \times 4180 \times \Delta T \times 60}{3.6 \times 1} = 2926 \times 10^3 \cdot \Delta T = 546 \times 10^3 - 506 \times 10^3 = 40 \times 10^3 \tag{35}$$

where, $F_V$ is the volume flow with the unit of m³/h, change it to the mass flow $F$ with the unit of kg/s,

$$F = \frac{F \cdot 1000}{3600} = \frac{F_V}{3.6} \text{ (kg/s)} \tag{36}$$

So,

$$\Delta T \approx 0.0137 \text{ ℃} \tag{37}$$

$$\delta t = \frac{\Delta T}{0.09} \times 60 \approx 9.1 \text{ s} \tag{38}$$

Similarly, for the collector power, we have,

$$\frac{F_V \cdot C \cdot \Delta T \cdot 60}{3.6 \cdot \tau} = \frac{39.1 \times 4180 \times \Delta T \times 60}{3.6 \times 1} = 2724 \times 10^3 \cdot \Delta T = 849 \times 10^3 - 812 \times 10^3 = 37 \times 10^3 \tag{39}$$

So,

$$\Delta T \approx 0.0136 \ °C \quad (40)$$

Because the inlet water temperature of collector is same as that of dummy load, the inlet water temperature rose 0.09 ℃ after 60 s,

$$\delta t = \frac{\Delta T}{0.09} \times 60 \approx 9.1 \ s \quad (41)$$

The analysis above is not very accurate because the temperature measurement is not exactly accurate, and the inlet water temperature increased gradually at just a nearly constant slope, but not a real constant slope.

TABLE 2. Power measurement results when the inlet water temperature is gradually increasing (No. 1), and results when the inlet water temperature is a constant (No. 2 ~ No. 6). All these data are measured when the pulse duration is 1s with the cathode voltage is -44 kV, the anode voltage is 21 kV, the main magnetic current is 50.4 A, the filament voltage is 33.4 V, and the filament current is 34 A.

| No. | Dummy Load Power / kW | Probable Error / % | Average Dummy Load Power / kW | Collector Power / kW | Average Collector Power / kW | Probable Error / % |
|---|---|---|---|---|---|---|
| 1 | 506 | | 506 | 812 | 812 | |
| 2 | 544 | | | 847 | | |
| 3 | 531 | | | 841 | | |
| 4 | 560 | 3.5 | 546 | 855 | 849 | 3.6 |
| 5 | 538 | | | 849 | | |
| 6 | 557 | | | 851 | | |

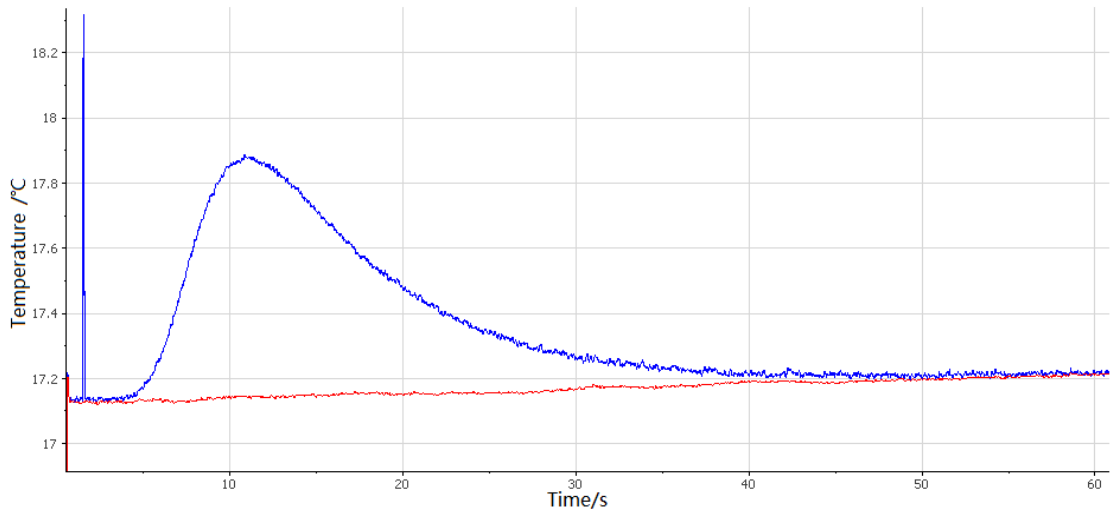

FIG. 8. The temperature of the dummy load outlet water and the inlet water along with the time in shot number 1. The blue curve indicates the outlet water temperature and the red curve indicates the inlet water temperature.

As shown in Fig. 9, when the inlet water temperature is decreasing gradually, the analysis is similar with that when the inlet water temperature increased gradually. We can get the correct power use formula (9), the definite integral is calculated as the green rectangle shown. But if we use formula (10) to calculate the power, the definite integral is calculated as the brown rectangle shown, the temperature difference is $\Delta T$ bigger than the correct value in every moment. So, the measurement result using formula (10) will be bigger than the correct result.

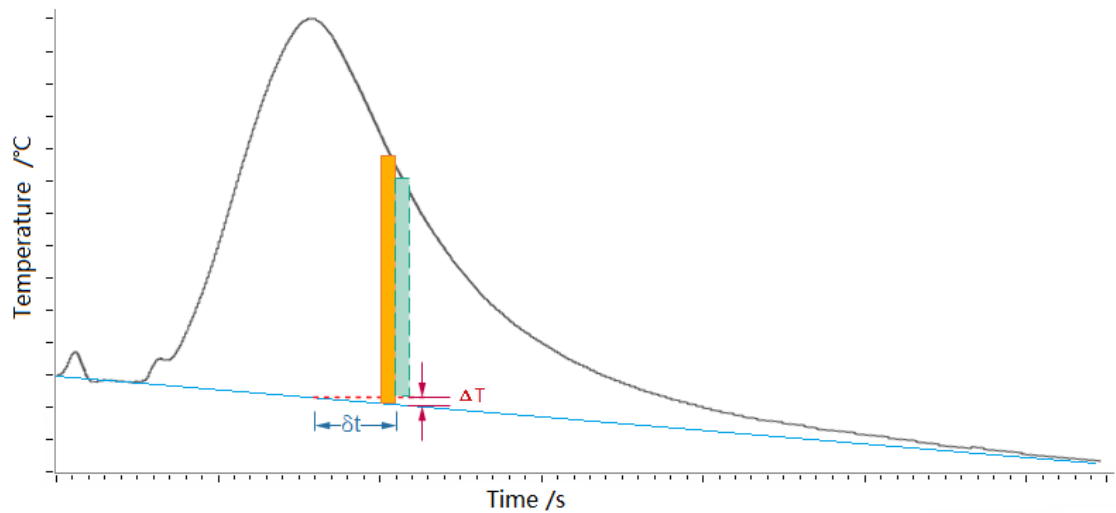

FIG. 9. The temperature of the outlet water and the inlet water along with the time when the inlet water temperature is gradually decreasing. The black curve indicates the outlet water temperature and the blue curve indicates the inlet water temperature.

## IV. TEST RESULTS

We have tested our gyrotrons and got the output power using calorimetric method. Up to now, the #1 Gycom gyrotron has demonstrated 900 kW output power (813 kW power is absorbed by dummy load) for 10 s with cathode voltage of -45.5 kV and anode voltage of 24 kV at about 48% electrical efficiency. Some related key signals are shown in Fig. 10. And the #1 Gycom gyrotron has demonstrated 644 kW output power (585 kW power is absorbed by dummy load) for 754 s[1] with cathode voltage of -44.5 kV and anode voltage of 24 kV at about 53% electrical efficiency.

The #2 CPI gyrotron has got about 499 kW output power (449 kW power is absorbed by aluminum tank dummy load and waveguide dummy load) for 79.66 s with cathode voltage of -57 kV and anode voltage of 21 kV at about 27.1% electrical efficiency and has got about 721 kW output power (649 kW power is absorbed by aluminum tank dummy load and waveguide dummy load) for 517 ms with cathode voltage of -59 kV and anode voltage of 22 kV at about 35.2% electrical efficiency. Some related signals are shown in Fig. 11 and Fig. 12. The long-pulse and high-power test is stopped by dummy load leaking problem.

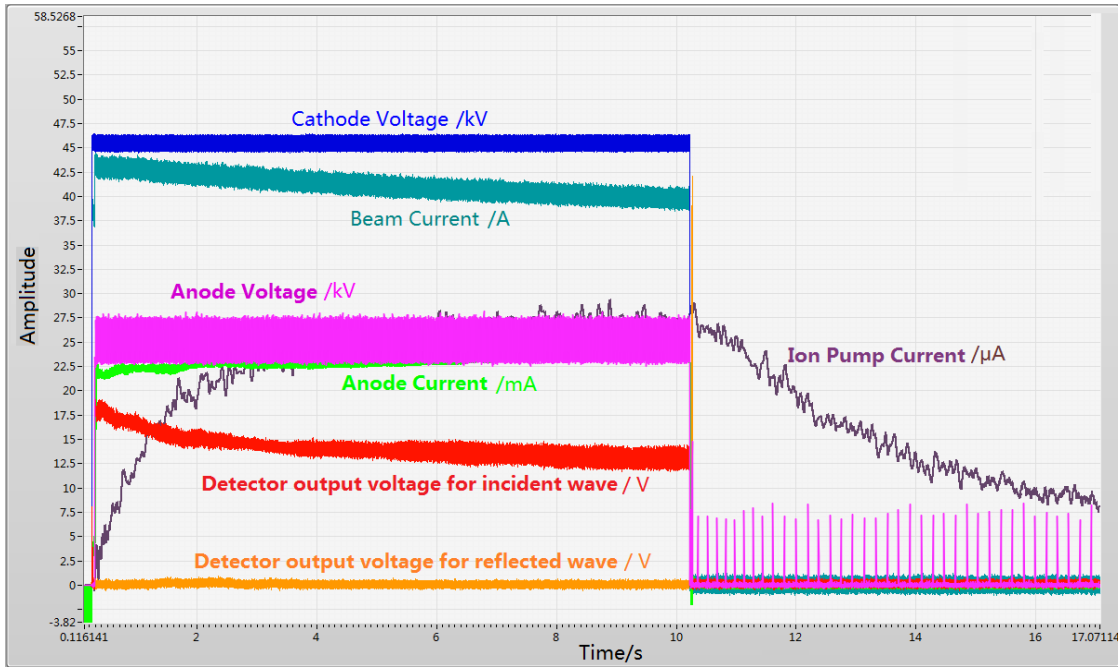

FIG. 10.  The key signals of Gycom gyrotron at 900 kW output power.

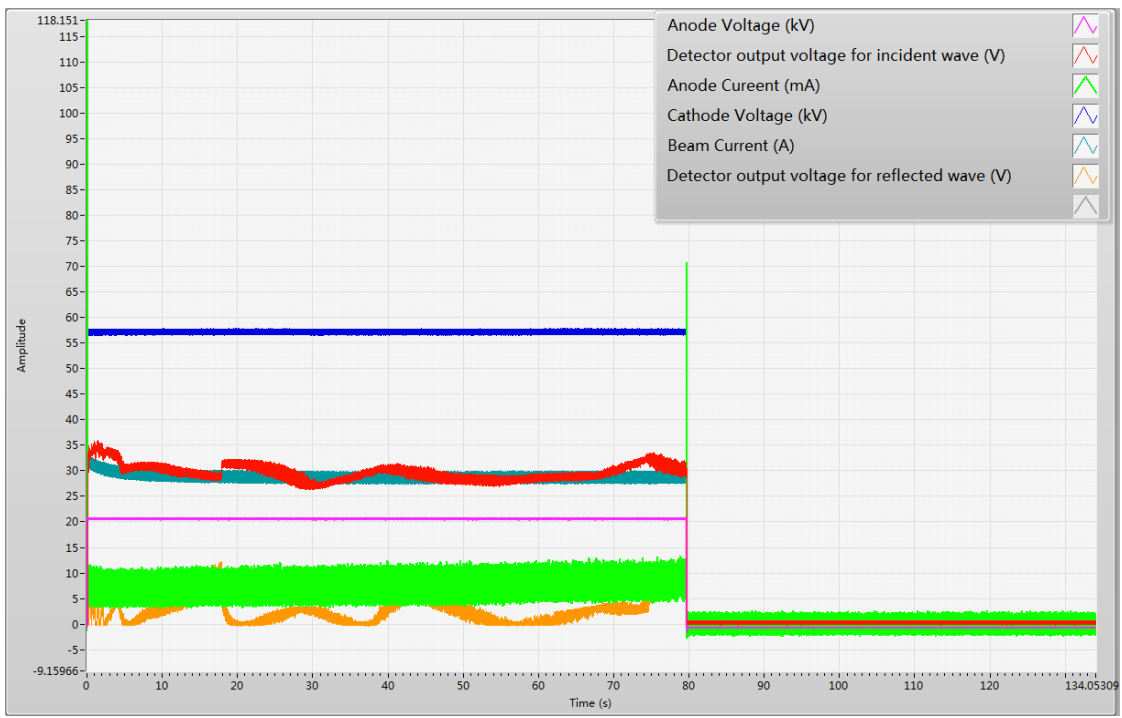

FIG. 11.  The key signals of CPI gyrotron at 499 kW output power.

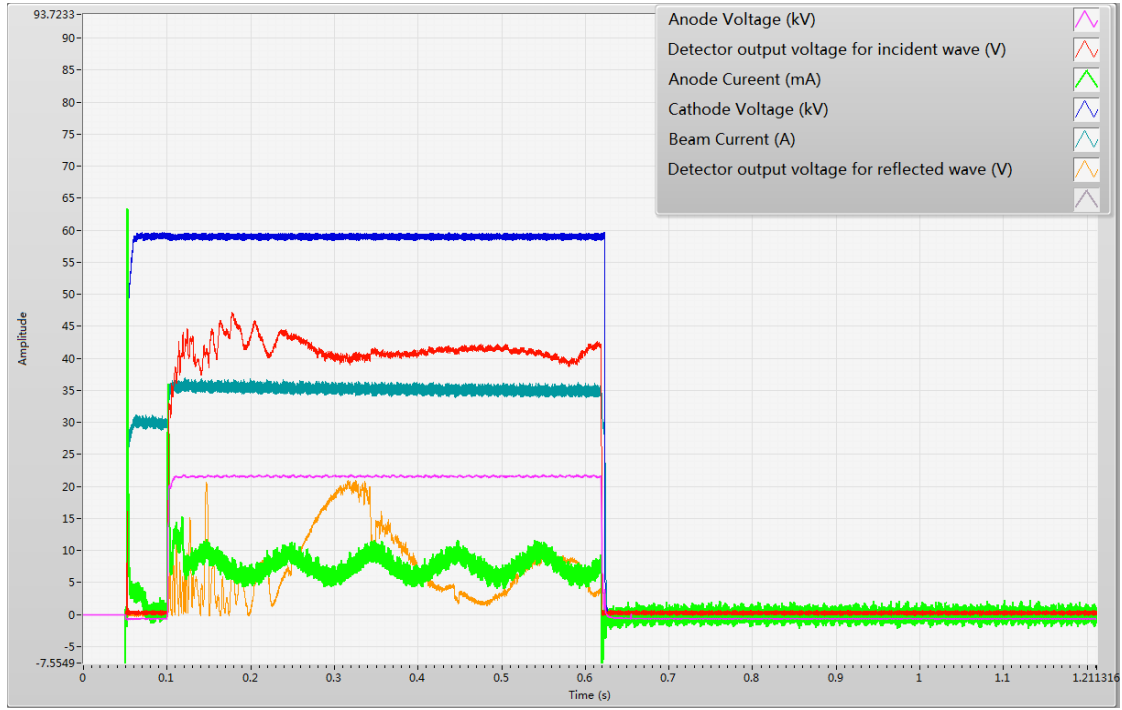

FIG. 12. The key signals of CPI gyrotron at 721 kW output power.

Except the dummy load and the collector, we measured the power of the other cooling circuits, such as Main Window, MOU etc.[15] The test results for #1 Gycom gyroton are shown in Table 3, and the probable calorimetric measurement error, water flow, and pressure drop for each cooling circuit are shown in Table 4. The power consumption is the sum of the cathode power and the anode power which are the product of voltage and current. As we can see, the sum of calorimetric power in all measured cooling circuits is approximately equal to the power consumption. It can prove that the calorimetric method is self-consistent, and the calorimetric measurement results are credible.

TABLE 3. The power of all measured cooling circuits for Gycom gyrotron. The power is in the unit of kW.

| No. | Pulse Duration (s) | Power Consumption | Dummy Load Power | Collector Power | MainWindow Power | Relief Window Power | Relief Load Power | Test Bench Power | MOU Power | Output RF Power | Sum of Calorimetric Power |
|---|---|---|---|---|---|---|---|---|---|---|---|
| 1 | 99.961 | 1005 | 306 | 644 | 0.62 | 1.07 | 4.5 | 5.45 | 23.2 | 334.65 | 984.84 |
| 2 | 99.981 | 1114.7 | 388 | 652.4 | 0.69 | 1.62 | 6.56 | 8.33 | 30.32 | 426.65 | 1087.92 |

TABLE 4. The probable calorimetric measurement error, water flow, and pressure drop for each cooling circuit.

| Cooling Cuicuit | Dummy Load | Collector | MainWindow | ReliefWindow | Relief Load | Test Bench | MOU |
|---|---|---|---|---|---|---|---|
| Probable Calorimetric Measurement Error (%) | 3.3 | 3.4 | 5.2 | 5.4 | 2.8 | 2.8 | 3.8 |
| Water Flow (m$^3$/h) | 42 | 39.6 | 2.3 | 2.2 | 5.2 | 5.1 | 3.4 |
| Pressure Drop (MPa) | 0.3 | 0.3 | 0.3 | 0.5 | 0.3 | 0.3 | 0.5 |

## V. CONCLUSION

Calorimetric measurement system has been established to measure the rf power generated by gyrotrons. The calorimetric measurements when the water temperature at the inlet is invariable or variable are discussed respectively. For getting more accurate results, we usually measure the dummy load power by keeping the inlet water temperature as a nearly constant. And the calorimetric measurement uncertainty has been discussed. The test results show that the calorimetric method is self-consistent and is credible for power measurements.

## ACKNOWLEDGMENTS

This work was supported by the National Magnetic Confinement Fusion Science Program of China (Grant No. 2011GB102000, No. 2015GB103000) and the Science Foundation of Institute of Plasma Physics Chinese Academy of Sciences (Grant No. Y45ETY230B). And we greatly appreciate the experts from CPI, GYCOM and GA for the cooperation in the development of ECRH project on EAST.